\documentclass[preprint,showpacs,preprintnumbers,amsmath,amssymb]{revtex4}

\usepackage{graphicx}% Include figure files
\usepackage{dcolumn}% Align table columns on decimal point
\usepackage{bm}% bold math

\newcommand{\bra}[1]{\langle #1|}
\newcommand{\ket}[1]{|#1\rangle}

\newcommand{\ketbra}[2]{| #1 \rangle \langle #2 |}
\begin{document}

%\preprint{APS/123-QED}

\title{Study  of the Distillability of  Werner States Using Entanglement Witnesses and Robust
Semidefinite Programs}% Force line breaks with \\

\author{Reinaldo O. Vianna$^\dagger$} 
\email{reinaldo@fisica.ufmg.br (permanent-address)}
\author{Andrew C. Doherty$^\dagger$}
\affiliation{$^{*\dagger}$School of Physical Sciences, The University of Queensland, 
Queensland 4072, Australia \\ 
$^*$ Departamento  de F\'{\i}sica - ICEX - Universidade Federal de Minas Gerais 
 \\ Av. Ant\^onio Carlos 6627 - Belo Horizonte - MG -  Brazil - 31270-901 }
\date{\today}

%\date{}

\begin{abstract}
We use Robust Semidefinite Programs and Entanglement Witnesses to
study the distillability of Werner states. We perform exact 
numerical calculations which show  2-undistillability 
in a region of the state space which was previously conjectured
to be undistillable. We also introduce 
bases which yield interesting expressions for the
{\em distillability witnesses} and for a tensor product
of Werner states with arbitrary number of copies.

\end{abstract}

\pacs{03.67.-a}% PACS, the Physics and Astronomy
\maketitle

\section{Introduction}

Maximally entangled states are the main resource in 
Quantum Information (QI) processing.
 Protocols like teleportation of quantum 
states \cite{teleporte} and entanglement based quantum cryptography \cite{cripto},
just to cite the two most emblematic,
work with perfect fidelity only when maximally entangled
bipartite states  are
available.( As maximally entangled bipartite states
are equivalent by local unitary transformations, they are
usually referred to as {\em singlets} in the jargon of QI.)
 On the other hand, even if one has a source of
perfect singlets, the ever present decoherence, due to 
interactions with the environment, degrades these states
to mixed form with reduced entanglement. Bennett  {\em et al.}
\cite{protocoloBennet} showed that this practical difficulty
could be circumvented by {\em distilling} singlets from
mixed states. This process involves only Local Quantum 
Operations and Classical Communication (LOCC) and, in
principle, is able to purify any mixed state to a singlet
form, given that an arbitrary supply of the former is
available. The process works at the expense of sacrificing
many of the mixed states in order to concentrate their
entanglement to a singlet. Horodecki  {\em et al.}
showed [4,5] that only states which violate
the Peres criterion \cite{Peres} can be distilled,
{\em i.e.}, the non-positivity 
of the partial transpose (NPT) is a necessary condition
for distillability.
 It was then realized
that there are entangled states which are not
directly useful in QI processing, they are said
to be {\em bound entangled} \cite{boundentanglement}, and the states with  positive
partial transpose (PPT) are of this kind. Nevertheless, these states can be {\em activated}
\cite{activation}, in the sense that, used in conjunction with
NPT states, they  can enhance the fidelity of teleportation.
Therefore we have two kinds of entanglement in nature, namely,
bound and free. The set of bound entangled states includes
all the PPT ones, but it is not known if there are NPT states
in this set. It was conjectured by DiVincenzo  {\em et al.} \cite{DiVincenzo}
and Dur  {\em et al.} \cite{Dur} that, in fact, there exist 
bound entangled NPT states. 

It can be shown that any  bipartite NPT state
can be transformed by LOCC to a Werner state \cite{Werner},
keeping the fidelity to the singlet.
The process is performed by  {\em twirling} (see
\cite{Dur}, for example) the states through the
action of bi-local unitary operations ($U\otimes U$).
Thus the study of distillability of arbitrary
bipartite states is reduced to the distillability
properties of Werner states.

Formally,  
a  bipartite state ($\rho \in {\cal B(H_A}\otimes {\cal H_B)}$) is distillable if and only if there exists
 a Schmidt rank two pure state ($\ket{\Psi} \in {\cal H_A}\otimes{\cal H_B}$), in the 
Hilbert space in which $\rho$ acts, such that
$\bra{\Psi}(\rho^{\otimes N})^{T_A}\ket{\Psi}$ is less than zero for some finite integer $N$
[5,8,9],
$T_A$ meaning partial transposition. When this condition is verified for some
$N$, $\rho$ is said to be $N$-distillable. In particular, all the bipartite entangled
states of the kind qubit-qudit ($2\times N$)  are 1-distillable [5,11].

In the same fashion that the entanglement of a state can
be decided by an {\em Entanglement Witness} operator \cite{Horodecki2223},
Kraus  {\em et al.} showed that the distillability also can be
decided by means of a kind of witness operator \cite{Kraus}.
In this paper, we show how to calculate these {\em 
distillability witnesses} using Robust Semidefinite
Programs (RSDP) \cite{RSDP} and apply it to study one- and 
two-distillability of Werner states. 
Starting with some definitions in section II, 
we revise the RSDP formalism in section III.
Sections IV and V present numerical results for
the distillability of Werner states in the
one- and two-copy cases.
In section VI, we derive some interesting expressions
for  the {\em distillability witness} and for a tensor
product of Werner states, with  arbitrary $N$, and 
then we conclude.

\section{Definitions}

The set of non-entangled (separable) states is convex and closed, therefore
it follows from the Separating Hyperplane  Theorem that there exists a linear 
functional ( {\em hyperplane}) that separates  an entangled state from
this set, this results in an Entanglement Witness \cite{Horodecki2223}.
Thus an EW ($W$) is a Hermitian operator with non-negative expectation value for
all the separable states, but which can have a negative expectation
value for an entangled state, in this case, the state is said to be
detected by the EW. An EW which can be written in the form
\begin{equation} 
W=P+Q_1^{T_A}+Q_2^{T_B}+\ldots+ Q_N^{T_Z},
\end{equation}
with $P$ and $Q_i$ positive operators, is said to be 
{\em decomposable}, and it is non-decomposable if it cannot
be put in this form. Only non-decomposable EWs can detect
PPT states. When the  EW (hyperplane) is tangent to the
separable set, it is said to be optimal (see 
\cite{FGSLB}, for example).

To be distillable [4,12], a bipartite state $\rho \in {\cal B(H_A}\otimes {\cal H_B)}$,
or a finite tensor product of it ($\rho^{\otimes N}$), must have a projection
which is NPT on a four-dimensional subspace, that is, given a Schmidt rank
two state,
\begin{equation}
\label{Eq2}
\ket{\Psi}=s_1\ket{e_1 f_1}+s_2\ket{e_2 f_2},
\end{equation}
where ${e_1,e_2}$ and ${f_1,f_2}$ are bases defining bi-dimensional subspaces
in ${\cal H_A}^{\otimes N}$ and  ${\cal H_B}^{\otimes N}$, respectively,
and $s_1$, $s_2$ are the Schmidt coefficients;
$\rho$ is distillable if the inequality
\begin{equation}
\label{desigualdade}
\bra{\Psi}(\rho^{\otimes N})^{T_A}\ket{\Psi}<0
\end{equation}
is satisfied for some arbitrary $\ket{\Psi}$, of the form of (\ref{Eq2}) , and some finite integer $N$.

We use the following parametrization of the Werner states \cite{Werner}:
\begin{equation}
\label{wern}
\rho_w=\frac{I_d+\beta  F_d}{d^2+d \beta},
\end{equation}
with $-1\leq \beta \leq 1$. $\rho_w$ is separable for
$\beta\geq-\frac{1}{d}$ and 1-distillable for $\beta<-\frac{1}{2}$.
$F_d$ is a swap operator for two qudits,
\begin{equation}
 F_d= \sum_{i,j=1}^{d}\ketbra{ij}{ji} ,
\end{equation}
and its partial transpose is the
bipartite maximally entangled state:
\begin{equation}
 P_d=\frac{1}{d} F_d^{T_A}= \frac{1}{d} \sum_{i,j=1}^{d} \ketbra{ii}{jj}.
\end{equation}
$I_d$ is the identity in the space of  the two qudits ($Tr(I_d)=d^2$).

Equivalently to the inequality (\ref{desigualdade}), Kraus  et.al. showed \cite{Kraus} that 
the distillability of an arbitrary 
 state $\rho$ (we will consider only the bipartite case $\rho \in {\cal B (H_A\otimes H_B)}$) can be decided 
 through the operator:
\begin{equation}
\label{Kraus-W}
W_N=P_2\otimes {(\rho^{T_A})}^{\otimes N},
\end{equation}
with $W_N$ acting in ${\cal (H}_{2A}\otimes {\cal H}_{2B})\otimes {\cal (H}_A  \otimes {\cal H_B)}^{\otimes N}$,  ${\cal H}_{2A}$ (${\cal H}_{2B}$)  being
the Hilbert space of a qubit belonging to A (B).
If $W_N$ is not an EW, then $\rho^{\otimes N}$ is N-distillable. If $W_N$ is a non-decomposable
EW, then  the  PPT  entangled state it detects ($\pi$) activates
 $\rho^{\otimes N}$, {\em i.e.}, $\rho^{\otimes N}\otimes \pi$ is 1-distillable.
 When $W_N$ is decomposable (and in the case of Werner states it happens to be a positive operator),
 then $\rho^{\otimes N}$ is undistillable and unactivable. When 
$W_N$ happens to be an EW, $W_N^{T_A}$ and $W_N^{T_B}$ are optimal EWs.

\section{Optimal Witnesses via Robust Semidefinite Programs }

Given a bipartite state $\rho \in {\cal B(H_A\otimes H_B)}$, we want to
determine its optimal entanglement witness $W_\rho$.
We will use the method introduced in \cite{FGSLB}, which we briefly describe
in the sequel.
Let $\Sigma$ be the set of separable states in ${\cal B}(H_A\otimes H_B)$.
We define the  following set (${\cal W}$) of entanglement witnesses:
\begin{equation}
{\cal W}=\{W\,\, /\,\, W\in {\cal B (H_A\otimes H_B)}; W^{\dagger}=W; 
Tr(W\sigma)\geq 0, \forall \sigma \in \Sigma; Tr(W)=1\}.
\end{equation}

$W_\rho$ is defined by:
\begin{equation}
\label{defW}
\min_{W\in{\cal W}} Tr(W\rho)=Tr(W_\rho \rho),
\end{equation}
and  can be determined through the following  
RSDP:
\begin{center}
\[\min_{W} Tr(W\rho),\]
\begin{equation}
\label{optimalW}
  subject\,\, to
\left\{
\begin{array}{l}
W^{\dagger}=W \\
 Tr(W)=1 \\
 \bra{\psi_A} W \ket{\psi_A}\geq 0 \,\,  
\forall  \ket{\psi_A} \in {\cal H_A}.
\end{array}
\right. 
\end{equation}
\end{center}
This is a NP-HARD problem and, in practice, we relax it to a
Semidefinite Program (SDP) by  taking a finite number ($n$) of pure states  ($\ket{\psi_A^i}$) to
represent the whole Hilbert space ${\cal H_A}$. Thus we replace an
infinite number of constraints ( {\em cf.} last line of (\ref{optimalW})) by
a finite set.
  If $dim({\cal H_A})$ is $d$, the kets $\ket{\psi_A^i}$
can be chosen as an uniformly distributed sample of unit complex vectors ($\vec{c_i}$) in 
${\cal C}^d$, and with infinite $n$  
this program would yield the exact witness $W_\rho$. In our calculations, we 
use an interior point algorithm to solve the SDP \cite{sedumi}.

In \cite{FGSLB2}, it was shown that  $W_\rho$ yields the random robustness ($Rr=-Tr(I)Tr(W_\rho \rho)$) of $\rho$, {\em i.e.},
the minimal amount of mixing with the identity necessary to wash out all the entanglement.
Thus, the state $\sigma=(\rho+Rr I/Tr(I))/(1+Rr)$ is in the border
between separable and entangled states.

Our main goal is to decide if the operator $W_N$ ({\em cf. (\ref{Kraus-W})}) is an
EW. We will do so by determining a state for which $W_N$ could be an optimal EW, in the sense
of (\ref{defW}). If we find such a state, we compare its optimal EW with the expression of
$W_N$, and this will tell if $W_N$ is or is not an EW.

\section{ONE-Copy Case }
We will  apply  the RSDP techniques to calculate optimal EWs ({\em cf.} (\ref{optimalW}) to
investigate the distillability properties of Werner states in the one-copy case.
We will show that the distillability  is related to the properties
of an EW for a certain family of PPT states.

We want to know if the Hermitian operator $W_1(\beta)\in{\cal B (H_A\otimes H_B)}$ defined by:
\begin{equation}
W_1(\beta)=P_2\otimes \rho_w^{T_A},
\end{equation}
is an entanglement witness.

We will show that, for $-\frac{1}{2}\leq\beta\leq -\frac{1}{3}$,
this operator is indeed a witness and, for $\beta=-\frac{1}{2}$, it is
the optimal witness ($W_\pi$) for a certain family of PPT entangled states
($\pi$).
Our numerical calculations will be restricted to qutrit
Werner states($d=3$), therefore $dim({\cal H_A})=dim({\cal H_B})=2\times d=6$.

We introduce the orthogonal basis:
\begin{equation}
\left\{
\begin{array}{l}
B_1=P_2\otimes P_3; \\
B_2=P_2\otimes (I_3-P_3);\\
B_3=(I_2-P_2)\otimes  P_3; \\
B_4=(I_2-P_2)\otimes (I_3-P_3).
\end{array}
\right.
\end{equation}
Note that ${B_i}/{Tr(B_i)}$ is an entangled state in 
${\cal B (H_A\otimes H_B)}$, and only $B_4$ is PPT.
In particular, $B_1$ is the maximally entangled state.

$W_1(\beta)$ can be recast as:
\begin{equation}
\label{Wx1-form}
W_1(\beta)=\frac{1}{d^2+\beta d}[(1+\beta d)B_1+B_2].
\end{equation}
In this basis, the states can be written as:
\begin{equation}
\begin{array}{l}
\rho=\sum_{i=1}^{4}p_i {B_i}/{Tr(B_i)},\\
\sum_{i=1}^{4}p_i=1,\,\,p_i\geq 0.
\end{array}
\end{equation}
Note that the state space is a 3-dimensional polytope defined by the
constraint $\sum_{i=1}^{4}p_i=1,\,\,p_i\geq 0 $.
Optimal witnesses for these states have the form:
\begin{equation}
\label{W1-form}
\begin{array}{l}
W=\sum_{i=1}^{4}c_i {B_i}/{Tr(B_i)},\\
\sum_{i=1}^{4}c_i=1,\,\,c_i\in\Re.
\end{array}
\end{equation}
Assuming  $W_1(\beta)$ is a witness for $-1\leq \beta \leq-\frac{1}{3}$,
we ask for the PPT state ($\pi$) it detects the most. It is done through the
following SDP:
\begin{center}
\[\min_{\pi} Tr[W_1(\beta) \pi],\] 
\begin{equation}
\label{receita1-pi}
subject\,\, to
\left\{
\begin{array}{l}
\pi^\dagger=\pi \\
\pi\geq0 \\
Tr(\pi)=1 \\
\pi^{T_A}\geq0

\end{array}
\right.
\end{equation}
\end{center}
We observe that the optimal solution ($\pi^*$)  of the above SDP is independent
of $\beta$ and minimizes this other SDP:
\begin{center}
\[\min p\]
\begin{equation}
\label{receita2-pi}
subject\,\, to
\left\{
\begin{array}{l}
\pi=(1-p)B_1/Tr(B_1)+pB_4/Tr(B_4)\\
\pi^\dagger=\pi \\
\pi\geq0 \\
Tr(\pi)=1 \\
\pi^{T_A}\geq0

\end{array}
\right.
\end{equation}
\end{center}
The optimal $p$ is $0.8571$, yielding the state $\pi^*$.
The optimal EW for this state, obtained by means of the RSDP (\ref{optimalW}), furnishes
$Tr(W_{\pi^*} \pi^*)=-0.0095$.

Comparing $Tr[W_1(\beta)\pi^*]$ with $Tr(W_{\pi^*}\pi^*)$, we observe that
%\begin{center}
\begin{equation}
%\left\{
%\begin{array}{l}
%Tr[W_1(\beta)\pi)]>Tr(W_\pi \pi),\,\, \forall -\frac{1}{2}<\beta\leq-\frac{1}{3};\\
%Tr[W_1(-\frac{1}{2})\pi)]=Tr(W_\pi \pi);\\
Tr[W_1(\beta)\pi^*)]<Tr(W_{\pi^*} \pi^*),\,\, \forall -1\leq \beta<-\frac{1}{2}.
%\end{array}
%\right.
\end{equation}
%\end{center}
These calculations are sufficient to show that $W_1(\beta)$ is not a
witness for $\beta <-\frac{1}{2}$, for it gives an expectation value
that is lower than that of the optimal EW.
On the other hand, for $ -\frac{1}{2}<\beta\leq-\frac{1}{3}$, 
we observe that $ Tr[W_1(\beta)\pi^*)]>Tr(W_{\pi^*}\pi^*)$,
and for $\beta=-\frac{1}{2}$, 
$Tr[W_1(-\frac{1}{2})\pi^*)]=Tr(W_{\pi^*} \pi^*)$.
With $W_{\pi^*}$ written in the
form (\ref{W1-form}), our calculations converge to 
$c_1=-\frac{1}{15}$, $c_2=\frac{16}{15}$ and $c_3=c_4=0$, which are
the parameters of $W_1(-\frac{1}{2})$. Therefore $W_1(-\frac{1}{2})$
is the optimal witness for $\pi^*$. This result can be obtained using
a large sample of random $\ket{\psi_A^i}$ ( {\em cf.} (\ref{optimalW})) or 
through  the following deterministic recipe.

Consider the  state ($\sigma$) defined by:
\begin{equation}
\label{receita-separaveis}
\sigma=\frac{\pi^*-Tr[W_1(-\frac{1}{2}) \pi^*]I}{1-(2d)^2Tr[W_1(-\frac{1}{2}) \pi^*]}.
\end{equation}
If $W_{\pi^*}$ and $W_1(-\frac{1}{2})$ coincide, they yield the random robustness
of  $\pi^*$ \cite{FGSLB2}, and $\sigma$ is a state in the border between
separable and entangled states ({\em viz.} Sec.III). Therefore $\sigma$ contains information
about the border of the separable set.
For each eigenvector ($\ket{\Psi_k}$) of $\sigma$, we form the state
$\rho_A^k=Tr_B(\ketbra{\Psi_k}{\Psi_k})$. Then we  use the eigenvectors
of the $\{\rho_A^k\}$ as a sample of states for the SDP (\ref{optimalW}).
In this case, this recipe yields 216 states, but just 24 already yield the
exact result.

Now we want to show that $W_1(\beta)$ is a witness for
$-\frac{1}{2}\leq\beta\leq-\frac{1}{3}$. For $\beta=-\frac{1}{2}$,
we know it is an optimal witness. Notice that $W_1(-\frac{1}{3})$
is a positive operator. Looking up (\ref{Wx1-form}), it is
easy to see that, for any state $\sigma$ and for $\beta_1<\beta_2$,
$Tr[W_1(\beta_1)\sigma)]\leq Tr[W_1(\beta_2)\sigma]$. In particular,
$Tr[W_1(-\frac{1}{2})\sigma]\leq Tr[W_1(\beta_2)\sigma]$.
If $\sigma$ is a separable state, $Tr[W_1(-\frac{1}{2})\sigma]\geq0$ and
therefore $Tr[W_1(\beta\geq-\frac{1}{2})\sigma]\geq 0$, showing it is
an entanglement witness.

All the calculations we have done can be understood more easily by
means of Figs.1 and 2.
Fig.1 shows a 2-dimensional
projection of a 3-dimensional plot of the state space. 
This picture is obtained as follows. We randomly draw $10^6$ states $\rho$.
Out of  each $\rho$, we build a border separable state ($\sigma$), and a state 
($\phi$) in the hyperplane defined by $W_1(\beta)$, namely:
\begin{equation}
\begin{array}{l}
\sigma=\frac{\rho-Tr(W_\rho \rho)I}{1-36 Tr(W_\rho \rho)};\\
\phi=\frac{\rho-Tr[W_1(\beta)\rho]I}{1-36 Tr[W_1(\beta)\rho]}.
\end{array}
\end{equation}
These states are rewritten in the zero trace basis $(I,G_1,G_2,G_3)$,
\begin{equation}
\begin{array}{l}
G_1=8B_1-B_2; \\
G_2=8B_3-B_4; \\
G_3=-3(B_1+B_2)+B_3+B_4;
\end{array}
\end{equation}
and their coefficients are plotted. We clearly see that the
planes $W_1(\beta)$ have a common axis, which is parallel to $G_3$.
In the picture, we can also see the state $\pi^*$ ( {\em cf.}
(\ref{receita1-pi}, \ref{receita2-pi})), which is in the 
plane $W_1(-\frac{1}{3})$. Notice that the plane
$W_1(-\frac{1}{2})$ is tangent to a face of the polytope defined
by the separable states. It can be clearly seen in Fig.2, which is
the 3-dimensional picture. Fig.2 also clearly illustrates the
concept of Optimal Entanglement Witness,  {\em i.e.}, a hyperplane
tangent to the separable set. Notice that in Fig.1, we have a
family of non-optimal EWs, the planes for $-\frac{1}{2}<
\beta\leq-\frac{1}{3}$. Therefore, these calculations show
that the Werner states are 1-undistillable for 
$-\frac{1}{2}\leq\beta\leq-\frac{1}{3}$.

 Of course the 1-undistillability of Werner states is not
a novelty [8,9]. The interesting result here
is the technique to decide if the Kraus-Lewenstein-Cirac operator
(\ref{Kraus-W}) \cite{Kraus} is an EW. The strategy was to compare
the candidate to EW with the optimal EW of a certain PPT state,
which can be obtained with arbitrary precision by means
of the RSDP (\ref{optimalW}), and to show that the candidate
operator converges to this EW. In this sense our calculations
are exact, leaving no room to doubt if the candidate operator
is or is not an EW. This technique extends straightforwardly to
higher dimensions. The other interesting result is that we
obtain a family of PPT entangled states,
 $\pi(p),\,\, 0.8571\leq p < 1$ ({\em cf. }(\ref{receita2-pi})), 
which activate the Werner states in the interval 
$-\frac{1}{2}\leq\beta<-\frac{1}{3}$, {\em i.e.}, $\rho_w\otimes\pi(p)$ is
1-distillable. In \cite{Vollbrecht}, similar  results
about the activation of Werner states were also obtained.

\begin{figure}
\includegraphics{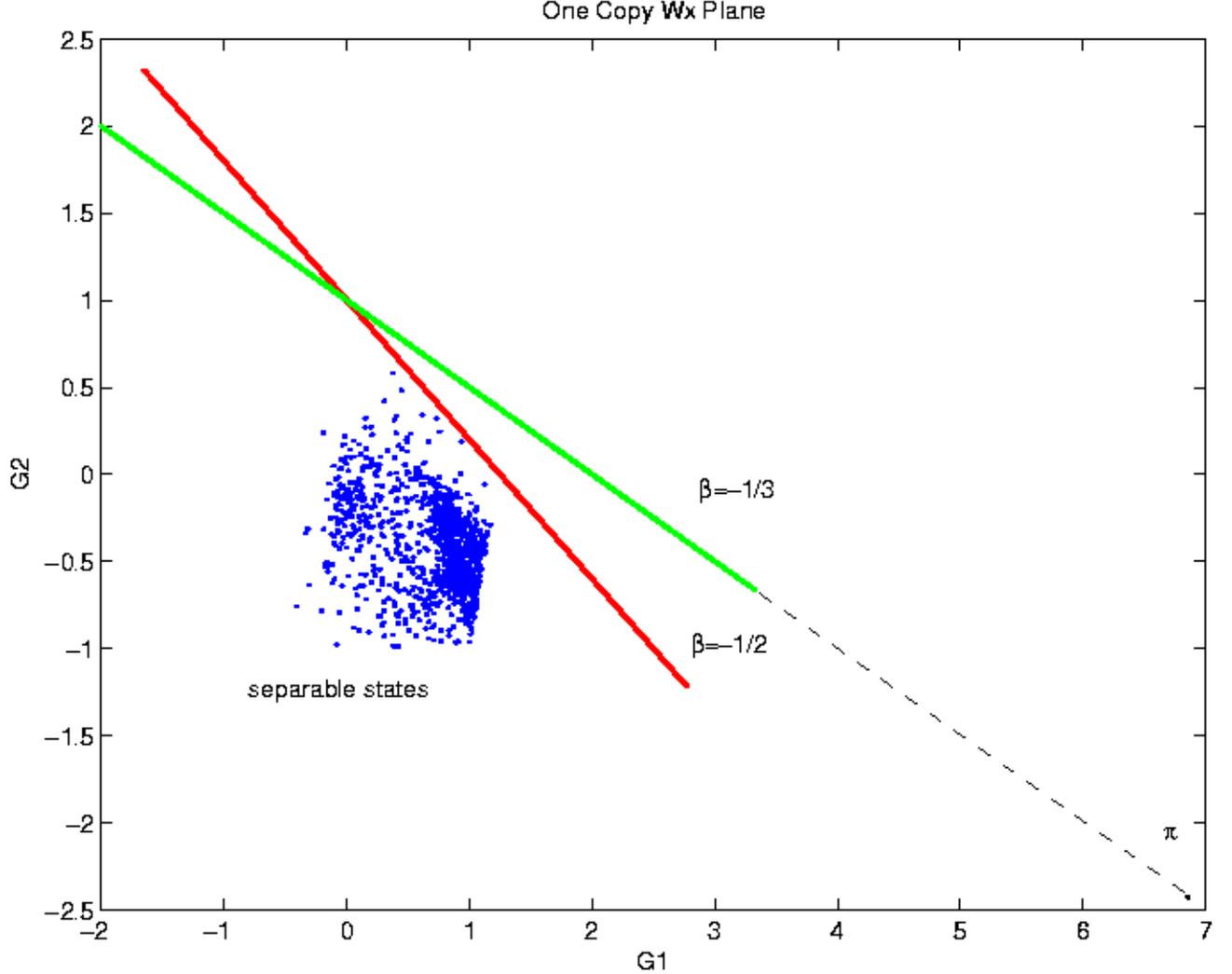}
\caption{  A projection of the state space showing the set of
separable states (just  border states - {\em cf.}(20)), the planes $W_1(\beta)$ and the PPT state $\pi$
for which $W_1(-\frac{1}{2})$ is the optimal witness. The planes
separate the state $\pi$ from the separable set.}
\end{figure}

\begin{figure}
\includegraphics{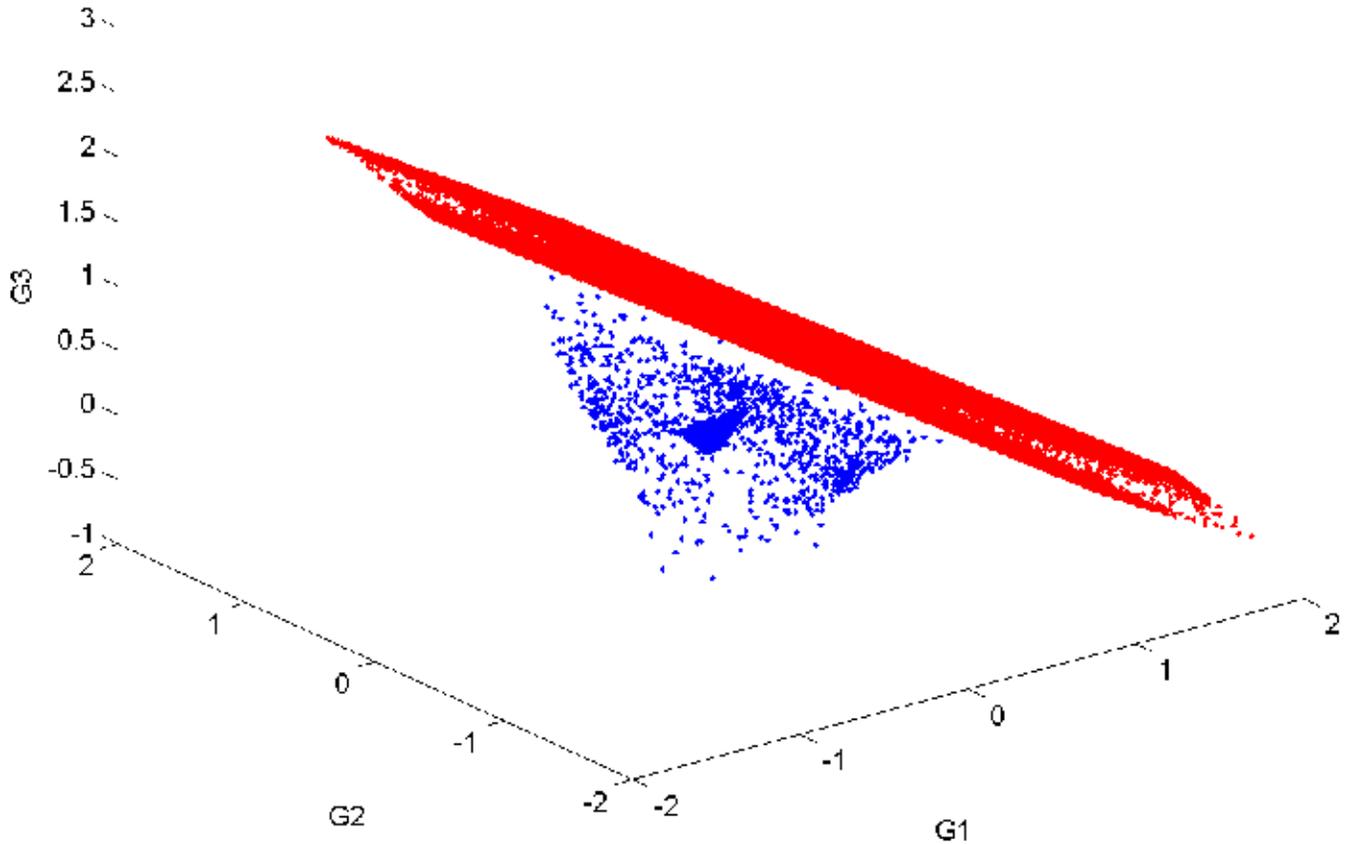}
\caption{  A three dimensional picture of the state space
showing  the set of separable states (just  border states - {\em cf.}(20)), which is a polytope, and
the plane $W_1(-\frac{1}{2})$ sitting on the polytope.
 $W_1(-\frac{1}{2})$ is an Optimal Entanglement Witness, 
 {\em i.e.}, a plane which is tangent to the separable set.}
\end{figure}

\section{TWO-Copy Case }

Now we will apply the techniques we have developed in the one-copy
case to study the distillability of Werner states in the two-copy
case. We will determine the optimal EW, by means of the RSDP (\ref{optimalW}),
for a family of PPT states. This will show that the Werner states
which are 1-undistillable are also 2-undistillable.

The calculations for the two-copy case mirror the one-copy case and
we arrive at  analogous conclusions. The orthogonal basis to expand the witnesses
and states is:
\begin{equation}
\left\{
\begin{array}{l}
B_1=P_2\otimes P_3\otimes P_3; \\
B_2=P_2\otimes [(I_3-P_3)\otimes P_3+ P_3\otimes (I_3-P_3)];\\
B_3=P_2\otimes (I_3-P_3)\otimes (I_3-P_3)]; \\
B_4=(I_2-P_2)\otimes P_3\otimes P_3; \\
B_5=(I_2-P_2)\otimes[(I_3-P_3)\otimes P_3+ P_3\otimes (I_3-P_3)];\\
B_6=(I_2-P_2)\otimes  (I_3-P_3)\otimes (I_3-P_3)]. 
\end{array}
\right.
\end{equation}
Again, if normalized, these are entangled states, with $B_1$ the
maximally entangled state, and $B_6$  the only PPT state. The
state space is a 5-dimensional polytope defined by:
\begin{equation}
\begin{array}{l}
\rho=\sum_{i=1}^{6}p_i B_i/Tr(B_i); \\
\sum_{i=1}^{6}p_i=1, \,\, p_i\geq 0.
\end{array}
\end{equation}

$W_2(\beta)=P_2\otimes (\rho_w^{T_A})^{\otimes 2}$ reads:
\begin{equation}
\label{Wx2-form}
W_2(\beta)=(\frac{1}{d^2+\beta d})^2[(1+\beta d)^2B_1+(1+\beta d)B_2+B_3].
\end{equation}

Using a SDP analogous to (\ref{receita1-pi}), we obtain that
the PPT state ($\pi$) most detected by $W_2(\beta)$ has the
coefficients $(0.0278,\, 0.2222,\, 0,\, 0,\, 0.0833,\, 0.6667)$.
We also obtain that $W_2(-\frac{1}{2})=W_\pi$ and $Tr(W_\pi \pi)=-0.0019$.
We can derive different families of PPT states for which $W_2(-\frac{1}{2})$
is optimal. A particularly interesting one reads:
\begin{equation}
\tilde{\pi}=(1-p)\frac{B_2}{Tr(B_2)}+p\frac{B_6}{Tr(B_6)},
\end{equation}
with $p$ exactly the same as in the one-copy case, namely,
0.8571.  For this state we have $Tr[W_2(-\frac{1}{2})\tilde{\pi}]=-0.0013$.

As our calculations show that $W_2(-\frac{1}{2})$ is an optimal witness,
$W_2(\beta)$ is an entanglement witness for $-\frac{1}{2}\leq \beta \leq 
-\frac{1}{3}$. Notice that these calculations are  exact and
show that qutrit Werner states are 2-undistillable in this interval
of $\beta$. On the other hand, they can be activated by the the
families of PPT entangled states detected by $W_2(-\frac{1}{2})$. 
The best results so far showed 2-undistillability analytically
[8,9] in the region
$-0.417\leq\beta\leq-\frac{1}{3}$, and provided numerical 
evidence  in  $-\frac{1}{2}\leq\beta\leq-0.417$.

We note that our calculations have been made in a laptop
with 1 GBytes of memory, and we used the symmetry of 
the Werner states to reduce the parameters in the optimization
problem. In  a larger computer, maybe the 3-copy case could
be handled, but the 4-copy case would need more than 
5 GBytes, just to load the basis set (\ref{base}).
One could try to explore the symmetry to reduce 
the matrices size, but in face of the constraints
in the Robust SDP ( {\em cf.} last line of \ref{optimalW}),
which is also the most memory consuming part of the calculations,
it  is far from trivial.

\section{The $W_N$ operator}
Although we are computationally limited to calculations for the 2-copy case,
we would like to understand the properties of the operator
$W_N$ when the number of copies ($N$) increases, hoping to shed light on the
general problem.  Indexing
the copies, we can write explicitly:
\begin{equation}
W_N=P_2\otimes \rho_{w1}^{T_A} \otimes \rho_{w2}^{T_A} \otimes \ldots \otimes \rho_{wN}^{T_A}.
\end{equation}
Note that it is normalized ($Tr(W_N)=1$) and tracing out the $N$th  copy yields 
the operator for $N-1$ copies:
\begin{equation}
Tr_N(W_N)=W_{N-1}.
\end{equation}
Now we  introduce a basis set which allows to write $W_N$ as a 
polynomial, generalizing  equations (\ref{Wx1-form}) and (\ref{Wx2-form}).

Define the following basis of  orthogonal projectors for the $N$-copy case:
\begin{equation}
\label{base}
\begin{array}{l}
B_1^N=P_2\otimes P_d^{\otimes N}; \\
B_{j+1}^N=P_2\otimes \frac{1}{(N-j)!j!}\sum_{i=1}^{N!}\hat{{\cal P}}_i[P_d^{\otimes(N-j)}\otimes(I_d-P_d)^{\otimes j}]; \\
B_{N+1}^N=P_2\otimes(I_d-P_d)^{\otimes N};
\end{array}
\end{equation}
with $B_{j+1}^N \in {\cal B}(H_A\otimes H_B)$. The ${\hat{\cal P}}_i$ 
  permute  the elements in the tensor product, yielding an expression
which is totally symmetric under exchange of any $P_d$ and $(I_d-P_d)$.
In this basis, $W_N$  has the diagonal representation:
\begin{equation}
\label{wx}
\begin{array}{l}
%W_N=\frac{1}{(d^2+d\beta)^N}\sum_{j=0}^{N}{(1+d\beta)^{N-j}B_{j+1}}
W_N=\sum_{j=0}^{N}\lambda_{j+1} B_{j+1}^N, \\
\lambda_{j+1}=\frac{(1+d\beta)^{N-j}}{(d^2+d\beta)^N}.
\end{array}
\end{equation}
We will show the correctness of this expression by induction.
Note first that 
\begin{equation}
Tr(B_{j+1}^N)=\binom{N}{j}(d^2 -1 )^j,
\end{equation}
where $\binom{N}{j}$ is the binomial coefficient. The basis for $N+1$ copies
is related to the $N$-copy basis through the recurrence relation:
\begin{equation}
\label{recorrencia}
B_{j+1}^N\otimes P_d + B_j^N\otimes (I_d-P_d)= B_{j+1}^{N+1}.
\end{equation}
If we normalize the basis (\ref{base}):
\begin{equation}
b_{j+1}^N\equiv\frac{B_{j+1}^N}{\binom{N}{j}(d^2-1)^j},
\end{equation}
we can rewrite (\ref{wx}) as:
\begin{equation}
\label{wx2}
W_N=\frac{1}{(d^2+d\beta)^N}\sum_{j=0}^{N} \binom{N}{j}(1+d\beta)^{N-j}(d^2-1)^j b^N_{j+1}.
\end{equation}
Now it is easy to see that the trace of (\ref{wx2}) is 1:
\begin{equation}
Tr(W_N)=\frac{\sum_{j=0}^{N} \binom{N}{j}(1+d\beta)^{N-j}(d^2-1)^j}{(d^2+d\beta)^N}=1.
\end{equation}
To finish the proof of the correctness of (\ref{wx}), we build $W_{N+1}$ by
appending $\rho_w$ to $W_N$:
\begin{equation}
\begin{array}{l}
W_{N+1}=W_N\otimes \rho_w^{T_A}= \\
\frac{1}{(d^2+d\beta)^{N+1}}\sum_{j=0}^{N}(1+d\beta)^{N-j}
B_{j+1}^N\otimes[(I_d-P_d) + (1+d\beta)P_d].
\end{array}
\end{equation}
Splitting  this  sum in two parts, and redefining the index in the 
second sum as $j+1=k$, we obtain:
\begin{equation}
\begin{array}{ll}
W_{N+1}= & \frac{1}{(d^2+d\beta)^{N+1}}\{\sum_{j=0}^N (1+d\beta)^{N+1-j}
B_{j+1}^N\otimes P_d  \\
          & +\sum_{k=1}^{N+1} (1+d\beta)^{N+1-k} B_k^N \otimes(I_d-P_d)\}.
\end{array}
\end{equation}
Writing out explicitly the terms for $j=0$ and $k=N+1$, we arrive at:
\begin{equation}
\begin{array}{ll}
W_{N+1}= & \frac{1}{(d^2+d\beta)^{N+1}}\{ (1+d\beta)^{N+1} B_1^N\otimes P_d  \\
& + \sum_{j=1}^N (1+d\beta)^{N+1-j}[B_{j+1}^N\otimes P_d + B_j^N \otimes(I_d-P_d)]  \\
& + B_{N+1}^N\otimes (I_d-P_d)\}.
\end{array}
\end{equation}
Finally, using the recurrence relation (\ref{recorrencia}) and the
basis definition (\ref{base}), we obtain the desired result:
\begin{equation}
\begin{array}{r}
W_{N+1}=\frac{1}{(d^2+d\beta)^{N+1}}\{ (1+d\beta)^{N+1} B_1^{N+1} +
\sum_{j=1}^N (1+d\beta)^{N+1-j} B_{j+1}^{N+1} + B_{N+2}^{N+1}\}= \\
\frac{1}{(d^2+d\beta)^{N+1}}\sum_{j=0}^{N+1}(1+d\beta)^{N+1-j} B_{j+1}^{N+1}.
\end{array}
\end{equation}

Once the correctness of (\ref{wx}) is proved, we highlight an interesting property
of the $W_N$ operator for Werner states.
If $|1+d\beta|<|d^2+d\beta|$, then all the eigenvalues of $W_N(\beta)$ go to
zero when $N$ tends to infinity, for any $-1\leq \beta \leq 1$.
This is an expected property. If $W_N(-\frac{1}{2})$ is an EW, for some $N$, 
it is necessarily optimal, and if properly normalized, it furnishes the
random robustness ($Rr$) for a family of entangled states ($\pi$), {\em i.e.},
$Rr=-(2d)^{2N} Tr(W_N(-\frac{1}{2})\pi)$ \cite{FGSLB2}, and we see
that the entanglement, as measured by the random robustness, increases
with $N$.

We can also  work out an expression analogous to 
(\ref{wx}) for a tensor product of Werner states:
\begin{equation}
\rho_w^{\otimes N}=\frac{(I_d+\beta F_d)^{\otimes N}}{(d^2+d\beta)^N}.
\end{equation}

We construct the following basis set, which has the same
algebraic structure of (\ref{base}), although it is not
orthogonal and only the last element is a positive operator:
\begin{equation}
\begin{array}{l}
 A_1^N=f_d^{\otimes N};\\
 A_{j+1}^N=\hat{S}[f_d^{\otimes(N-j)}\otimes (I_d-f_d)^{\otimes j}];\\
A_{N+1}^N=(I_d-f_d)^{\otimes N};
\end{array}
\end{equation}
with $A_{j+1}^N \in {\cal B}(H_A\otimes H_B)$, and $f_d\equiv F_d/d$.
Note that this basis is obtained by discarding $P_2$ in (\ref{base}) and
by taking the partial transpose of $P_d$.
In this basis, $\rho_w^{\otimes N}$ reads:
\begin{equation}
\rho^{\otimes N}=\frac{1}{(d^2+d\beta)^N}\sum_{j=0}^{N}(1+d\beta)^{N-j}A_{j+1}^N.
\end{equation}

Note that $A_{N+1}^N$ is a fully separable positive operator. In particular,
if we take $\beta=-\frac{1}{d}$ then
\begin{equation}
\rho^{\otimes N}=\frac{A_{N+1}^N}{Tr(A_{N+1}^N)}=\frac{A_{N+1}^N}{(d^2-1)^N}.
\end{equation}
Normalizing the  basis:
\begin{equation}
a_{j+1}^N\equiv \frac{A_{j+1}^N}{Tr(A_{j+1}^N)}=\frac{A_{j+1}^N}{\binom{N}{j}(d^2-1)^{j}},
\end{equation}
$\rho^{\otimes N}$ reads:
\begin{equation}
\label{rhow}
\rho^{\otimes N}=\frac{1}{(d^2+d\beta)^N}\sum_{j=0}^{N}\binom{N}{j}(1+d\beta)^{N-j}(d^2-1)^{j}a_{j+1}^N.
\end{equation}
Then, for $\beta=-\frac{1}{d}$ , $\rho^{\otimes N}=a_{N+1}$. When we take $N$ to infinity,
the binomial coefficients in (\ref{rhow}) with $j\neq0$ and $j\neq  N$ dominate the sum,
but nothing special seems to occur.

\section{Conclusions }
We have done exact numerical calculations which show the 
2-undistillability of qutrit Werner states in the
region $-\frac{1}{2}\leq\beta\leq-\frac{1}{3}$. 
We have shown that $W_N(-\frac{1}{2})$ is an optimal entanglement
witness (for $N=1,2$), and this is our certificate of 1- and 2-undistillability.
We have derived families of PPT entangled states which activate
the 1- and 2-undistillable Werner states.
%In the 1-copy case, we concluded that  $W_1(-\frac{1}{2})$ is 
%tangent and parallel to a face of the separable set, which happens
%to be a polytope.
 We have introduced a basis of orthogonal projectors to 
expand $W_N(\beta)$, which shows that this operator is a 
polynomial in $(1+d\beta)$, and it acts in a state space
which is a polytope. 
The eigenvalues of  $W_N(\beta)$ tend to zero, when we consider
infinite many copies of Werner states and it is a property related
to the random robustness of the states it detects. We have also introduced
a basis which provides a simple polynomial expression for
a tensor product of Werner states.

ACKNOWLEDGMENTS: ROV is grateful to Paulo H. Souto Ribeiro
and Fernando G.S.L. Brand\~ao for their criticism and suggestions
on the manuscript. ROV also  acknowledges the financial support provided
by the Brazilian agencies CAPES, FAPEMIG, CNPQ and Instituto do Mil\^enio 
de Informa\c c\~ao Qu\^antica MCT/CNPq.


\begin{thebibliography}{99}
\bibitem{teleporte} C.H. Bennett, G. Brassard, C. Crepeau, R. Jozsa, A. Peres,
W.K. Wootters, Phys. Rev. Lett. {\bf 70}, 1895 (1993).
\bibitem{cripto} A.K. Ekert, Phys. Rev. Lett. {\bf 67}, 661 (1991).
\bibitem{protocoloBennet} C.H. Bennett, G. Brassard, S. Popescu, B. Schumacher,
J.A. Smolin, W.K. Wootters, Phys. Rev. Lett. {\bf 76}, 722 (1996).
\bibitem{boundentanglement} M. Horodecki, P. Horodecki, R. Horodecki, Phys. Rev.
Lett. {\bf 80}, 5239 (1998).
\bibitem{reducao} M.Horodecki, P.Horodecki, Phys. Rev. A
{\bf 59}, 4206 (1999).
\bibitem{Peres} A. Peres, Phys. Rev. Lett. {\bf 77}, 1413 (1996).
\bibitem{activation} P. Horodecki, M. Horodecki, R. Horodecki, Phys. Rev. Lett. 
{\bf 82}, 1056 (1999).
\bibitem{DiVincenzo} D.P. DiVincenzo, P.W. Shor, J.A. Smolin, B.M. Terhal,
A.V. Thapliyal, Phys. Rev. A {\bf 61}, 062312 (2000).
\bibitem{Dur} W.D\"ur, J.I.Cirac, M.Lewenstein, D.Buss, Phys. Rev. A 
{\bf 61}, 062313 (2000).
\bibitem{Werner} R.F. Werner, Phys. Rev. A {\bf 40}, 4277 (1989).
\bibitem{Horodecki2223}  M. Horodecki, P. Horodecki, R. Horodecki, Phys. Lett. A  {\bf 223},
1 (1996).
\bibitem{Kraus} B. Kraus, M. Lewenstein, J.I. Cirac, Phys. Rev. A {\bf 65}, 042327 (2002).
\bibitem{RSDP} see, for example, A. Ben Tal, A. Nemirovski, Math. Operat. Res. {\bf 23},
769 (1998).
\bibitem{FGSLB}F.G.S.L Brand\~ao and R.O. Vianna, Phys. Rev. Lett.
{\bf 93}, 220503 (2004).
\bibitem{sedumi} J.F. Sturm, Optimizations Methods and Software {\bf 11-12},
625 (1999), SEDUMI: \begin{tt}http://fewcal.kub.nl/sturm/softwares/sedumi.html\end{tt};
J. L{\"o}fberg,
 {YALMIP} : A Toolbox for Modeling and Optimization in {MATLAB},
 Proceedings of the {CACSD} Conference,
 2004,
Taipei, Taiwan,
 \begin{tt}http://control.ee.ethz.ch/\~{}joloef/yalmip.php\end{tt};
   S. Boyd, L. Vandenberghe, {\em Convex Optimization}, 
(Cambridge University Press, Cambridge, 2000).
%\bibitem{Watrous} J.Watrous, Phys. Rev. Lett. {\bf 93}, 010502 (2004).
\bibitem{FGSLB2} F.G.S.L. Brand\~ao, R.O.Vianna, Int. J. Q. Inf. 
{\bf 4}, 331 (2006).
\bibitem{Vollbrecht} K.G.H. Vollbrecht, M.M. Wolf, Phys. Rev. Lett.
{\bf 88}, 247901 (2002).

\end{thebibliography}
\end{document}